# Metallic bonds become molecular-like in atomic-sized devices

Harsh Deep Chopra[*], J. N. Armstrong and Susan Z. Hua[*]

*Laboratory for Quantum Devices, Materials Program, Mechanical and Aerospace Engineering Department, The State University of New York at Buffalo, Buffalo, NY 14260, USA*

**Abstract:** Covalent molecules are characterized by directed bonds, which provide stability-of-form to the molecule's relative atomic positions. In contrast, bulk metals are characterized by delocalized bonds, where a large number of resonance structures ensure their high stability. However, reduced to atomic dimensions, metallic arrangements become increasingly vulnerable to disruptive entropic fluctuations. Using the smallest possible device, namely, a single atom held between two atomically sharp probes, force to rupture single-atom bridges was measured with pico-level resolution, using gold and silver. Remarkably, measured forces are found to be a precise vector sum (directional bonding) of cohesive forces between the central and adjacently coordinated atoms. Over three to four times stronger than bulk, the directional bonds provide high configurational stability to atomic-sized metallic devices, just as delocalization-induced resonance stabilization is the emergent response of bulk metals. Results open new opportunity for molecular electronics without complications arising from metal/molecule interfaces.

Covalent molecules are characterized by directed bonds, which provide stability-of-form to the molecule's relative atomic positions. In contrast, the hallmark of bulk metals is the metallic orbital, which permits unsynchronized resonance of electron pair bonds from one atomic site to another.[1-4] It allows for a large number of resonance structures, which in turn, provide high stabilization and an unlimited capacity for the 'unsaturated metallic molecule' to add more atoms.[1-6] It also lends metals their characteristic properties such as high electrical and thermal conductivity, strength, ease of atomic displacement from normal lattice sites (deformation), etc.[5-7] While delocalization-induced resonance stabilization explains bulk behavior, metallic systems reduced to atomic dimensions become increasingly susceptible to entropic fluctuations. With increasing drive towards miniaturization, it is therefore important to understand factors governing the inherent stability of atomic sized metallic devices. In this context, a single atom held between two atomically sharp probes is the smallest 'sample' that can be probed for its specific elemental properties. Practically, it is the ultimate physical limit in device miniaturization.[8-13] It is also the first or last atomic configuration to occur in many phenomena (e.g., during coalescence of islands in thin film growth, as starting point or leading edge of a crack, as contact points between adjoining surfaces in tribology or friction, etc.). In the present study the force to rupture single-atom gold and silver bridges was measured with pico-level resolution. Measurements reveal three distinct morphologies in which the central atom is coordinated with one, two or three adjacent atoms on either side of the bridge. The forces to rupture these morphologies are found to be a precise vector sum of forces between the central and adjacently coordinated atoms, revealing the molecular-like bonds in metals with extremely starved coordination. Three to four times stronger compared to bulk, the directional bonds impart high configurational stability to atomic-sized devices.

The experimental procedure is described in detail in previous publications.[14-16] Briefly, gold and silver films (200 nm thick) were magnetron sputtered on silicon substrates and cantilever tips at 30 W in argon partial pressure of 3 mtorr. The purity of the sputtering targets was 99.999%. The base pressure of the UHV chamber was $10^{-8}$-$10^{-9}$ torr. During deposition the cantilevers were rotated relative to the sputtering gun to enhance film uniformity in the vicinity of the tip. The AFM based probe to make stable single-atom bridges and measure pico-level forces or deformations induced by mechanical excitations is based on a dual piezo approach.[14, 15, 17] With this probe the coarse piezo is used to initially close the gap between the tip and the substrate. Then a fine piezo is used for desired experiments, which can position the substrate in x-y-z directions. The probe is capable of forming a stable single-atom bridge between an atomically sharp probe and substrate without any feedback loop; precisely deform it in increments of sub-atomic distances (noise band is 5 pm peak-to-peak and its center line can be shifted by a minimum step of 4 pm); measure bond-bond strength at pN level; simultaneously measure conductance across the bridges; and controllably change the size of the sample virtually atom-by-atom. The system has enough steadiness and stable single-atom contacts can be readily formed without the need for any feedback. Measurements were made at room temperature in an inert atmosphere chamber. The assembly is enclosed in an acoustic chamber and a Faraday cage. For enhanced stability the probe assembly is mounted on a three-stage vibration isolation system to minimize the destabilizing effects of mechanical vibrations. Conductance traces were recorded at a bias voltage of 50 mV. A range of cantilever spring constants were used (11-80 N/m). The cantilevers were precisely calibrated using reference cantilevers available from Veeco Probes (Force Calibration Cantilevers CLFC-NOBO). The piezo was extended or retracted at the rate of 5 nm/s.

A schematic of the experimental setup to form highly stable atomic sized bridges and measure forces to rupture them is shown in Fig. 1(a). The experimental procedure is given in previous publications[14-16] and is briefly summarized in the Methods section. Figure 1(b) shows an example of the simultaneously measured force and conductance as a piezo is being controllably retracted to break an initially large gold bridge down to a single-atom bridge followed by its rupture. From such traces the force $F$ to rupture a single-atom bridge can be measured while the simultaneously measured quantized conductance across the bridge unambiguously ascertains the existence of a single atom at the narrowest point of the constriction. The schematic in Fig. 1(a) also shows various morphologies of a single-atom bridge corresponding to a coordination of one, two or three atoms. Hybrid or mixed morphologies can also occur, such as the coordination of two atoms on one side and three on the other side of the single atom, as shown in Fig. 1(a). However, the force to rupture the bridge is uniquely defined by the number of broken bonds (1 or 2 or 3).

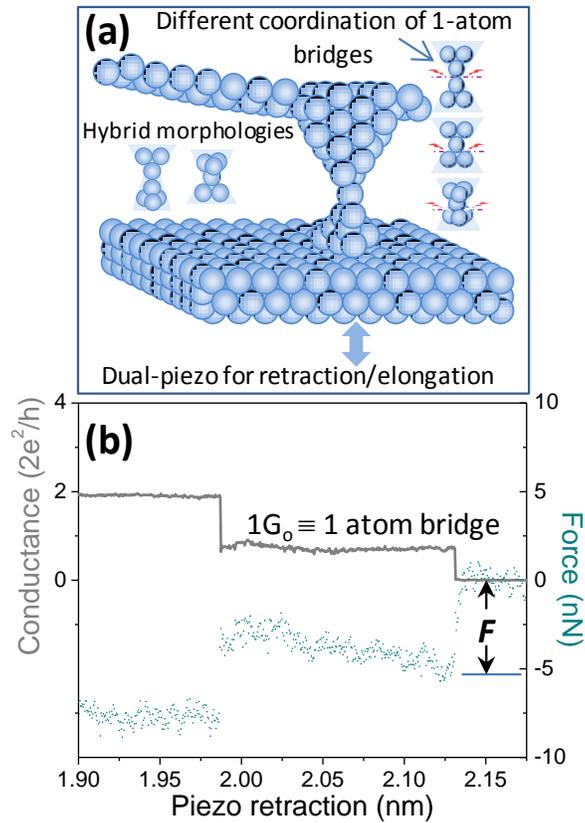

**Fig. 1.** Measurement of force to rupture a single-atom bridge formed between a film and an atomically sharp cantilever tip coated with the same metal. (a) Schematic showing that the central single-atom can have coordination of one, two or three on either side. The setup uses a dual-piezo assembly to form and break the bridges, as explained in the Methods. (b) Simultaneously measured force and conductance versus piezo retraction across atomic sized gold bridges. The last conductance plateau at $1G_o$ defines a single atom gold bridge. The magnitude of force $F$ to rupture the single-atom bridge is also shown.

The distribution of measured forces to break single-atom Au bridges over repeated experiments is shown in Fig. 2(a). The abscissa in Fig. 2(a) is the simultaneously measured conductance, which can have arbitrary values less than or equal to $1G_o$ depending on the probability of electron transmission across the bridge. Here, $1G_o$ ($= 2e^2/h$) is the quantum of conductance equal to $7.748091 \times 10^{-5}$ siemens (~$1/12{,}906$ $\Omega^{-1}$); $e$ is the quantum of charge and $h$ is Planck's constant.[18-20] The conductance of a 1-atom gold (and silver) bridge cannot exceed $1G_o$ since monovalent gold (and silver) has only a single available channel whose maximum conductance is $1G_o$. This is an inviolable condition of quantum mechanics and permits single-atom bridges (i.e., a bridge where the narrowest constriction is the diameter of a single-atom) to be distinguished from larger diameter bridges; this is discussed in detail in a previous publication.[14] Figure 2(a) shows that over repeated experiments the breaking force becomes clustered around three mean values (2.06 nN, 3.42 nN and 4.86 nN), as seen statistically in Fig. 2(b). The observed clustering corresponds to different morphologies of the single-atom bridge where the central atom is coordinated with either one, two or three atoms, as shown schematically in Fig. 2(c). As noted earlier in Fig. 1(a), hybrid (or asymmetrical) morphologies can also exist. However, the force to rupture the bridge is uniquely determined by the number of broken bonds (either 1 or 2 or 3). *A priori* one might have assumed that the force to rupture two or three bonds would be integral multiples (or algebraic sum) of the force to break a single bond. However, it is clear from Fig. 2 that this bulk assumption does not hold. At first sight, the mean values of the measured forces to break bridges having different atomic coordination appear to be unrelated to each other. However, a closer analysis reveals a simple relationship. As illustrated in Fig. 2(c), the average value of force to rupture a single Au bond $F_b$ is ~2.06 nN. The force to rupture a bridge with coordination of two atoms ($\theta_{c,2} = 30°$) is $2F_b cos\theta_{c,2} = 3.56$ nN, as shown in the free-body force diagram in Fig. 2(c). This is in close agreement with the measured value of 3.42 nN in contrast to twice the force to rupture a single bond, $2F_b = 4.12$ nN. For a coordination of three bonds ($\theta_{c,3} = 35.26°$) the free body diagram in Fig. 2(c) indicates that the force to rupture the bridge would be $3F_b cos\theta_{c,2} = 5.04$ nN. This agrees well with the measured value of 4.86 nN, showing again that 'directed' behavior for cohesive forces ($F_{rupt} = nF_b cos\theta_{c,n}$) represents such bonding far better than 'non-directed' behavior ($F_{rupt} = nF_b$).

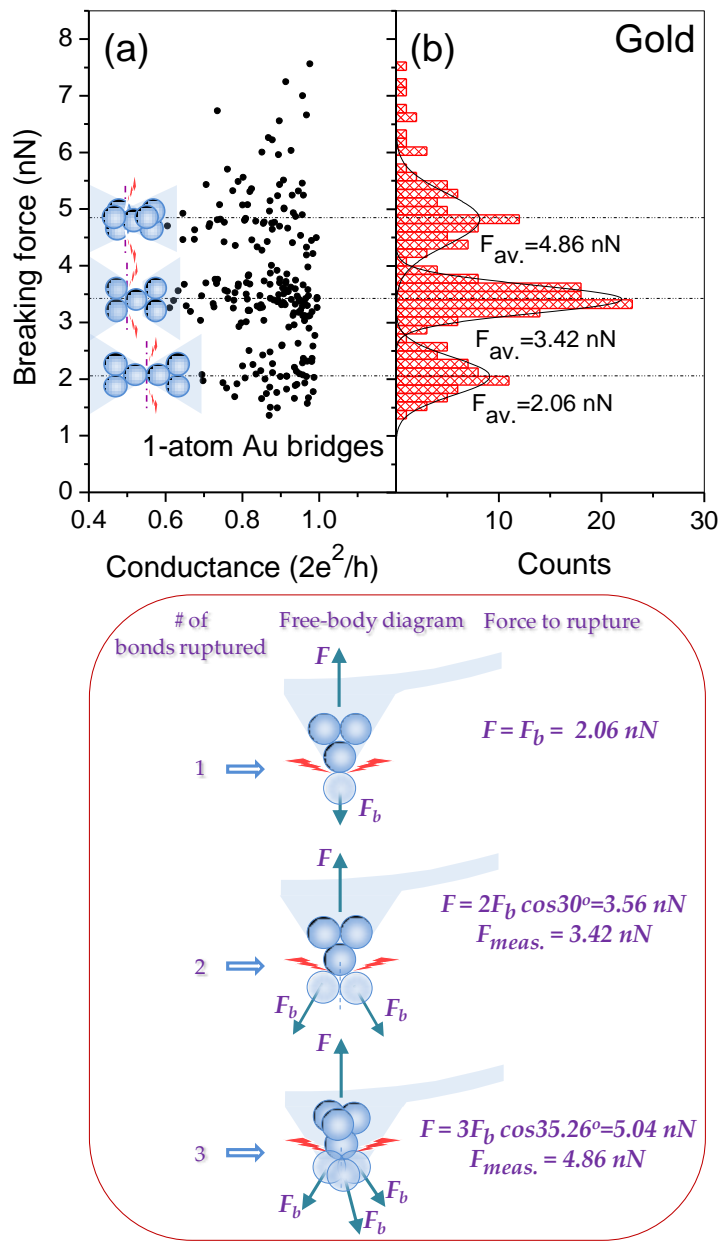

**Fig. 2**. Force to break single-atom gold bridges. (a) Distribution of measured forces over repeated experiments shows clustering of the data into three distinct bands, as shown statistically in (b). (C) Free-body force diagrams show that the force to break coordination of two and three atom bonding is a vector sum of forces between adjacent atoms.

Similar behavior is observed when single-atom silver bridges were ruptured, as shown in Fig. 3(a-b). For the case of silver, the mean value of the measured force to break a single bond $F_b$ is ~1.06 nN. Again the second peak (at 1.8 nN) corresponding to atomic coordination of two atoms closely follows the free-body vector relationship, $2F_b cos30^o$ or 1.84 nN, and the directional nature of the bonding is self-evident. In contrast to gold, silver rarely formed single-atom bridges with a coordination of three atoms (only one instance was found). Figure 2(b) and 3(b) also shows the relative probability to form bridges with different atomic coordination. For example, gold is twice as likely to form a single-atom bridge having atomic coordination of 2 versus 1 or 3. In contrast, Ag has a much higher probability to form atomic coordination of 1 instead of 2. Also note that several factors might contribute to the observed dispersion of forces around each peak in Figs. 2 and 3. Magnitude of the measured forces would vary in the presence of an imperfect stacking sequence of close packed planes (stacking faults) across the bridge. For example, a hexagonal stacking sequence along the bridge would cause the cohesive forces to differ relative to the normal stacking in face centered cubic crystals. Hybrid bridge morphologies could have a similar effect.

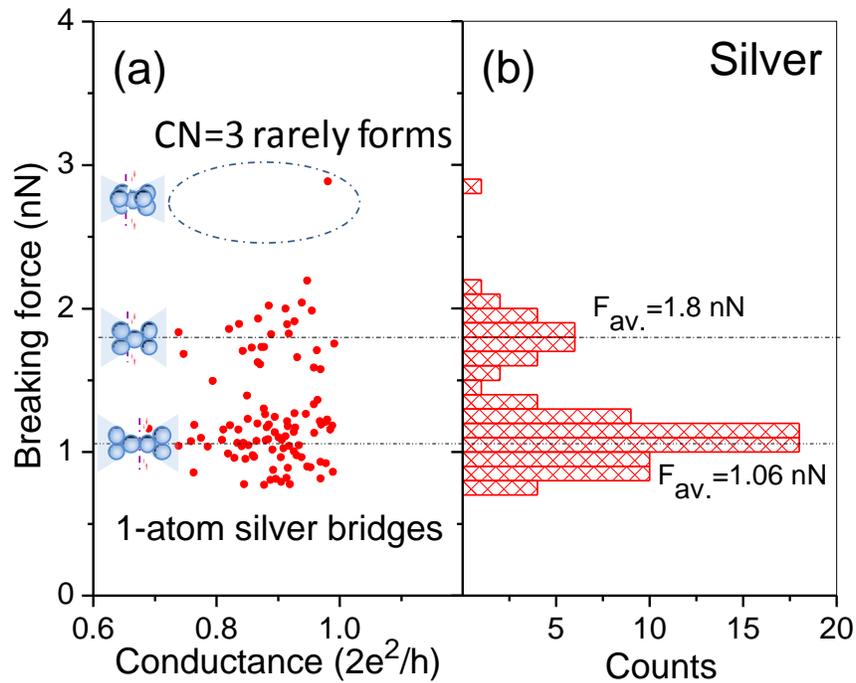

**Fig. 3**. Force to break single-atom silver bridges. (a) Distribution of measured forces over repeated experiments shows clustering of the data into two distinct bands, as shown statistically in (b). Notice that for silver a coordination of three atoms rarely forms.

It is clear that when metallic atoms are placed in configurations of lowest possible coordination, bonding becomes molecular-like. With increasing coordination experiments show that the molecular-like bonds rapidly evolve into metallic bonds (not shown). Acute reduction in atomic coordination along with confinement effects are well known to cause properties to deviate strongly from the bulk. For example, conductance across the bridge becomes quantized,[14, 21] spintronics effects arise from quantum domain walls instead of classical scattering,[9, 22, 23] composition effects become enhanced,[24, 25] and mechanical properties approach ideal values.[15, 16] Whereas quantization effects are due to confinement, enhanced mechanical properties are associated with changes in cohesive forces. Indeed, bond stiffening and contraction at reduced atomic coordination was predicted by Pauling and others[26, 27] and has recently been experimentally validated in terms of enhanced modulus that is 2-4 times higher than in bulk[15] and strength approaching theoretical values in the limit of a single atom bridge.[16] Using cohesive energy for gold (3.81 eV/atom) and coordination of 12, the force to break a single Au-Au bond in the bulk is 0.53 nN (assuming a range of force of ~0.096 nm, which is a third of Au-Au bond length of 0.288 nm); estimates using modulus of bulk gold yield a value of 0.8-0.9 nN whereas density functional theory calculations give an estimated value of no more than 0.7 nN.[28] In other words, the measured value of 2.06 nN in Fig. 2 to break a single Au-Au bond reflects significant strengthening, as predicted by Pauling. Previously, a somewhat lower value of ~1.5 nN was reported to break apart a single gold atom at room temperature.[29] Initially it was attributed to rupture of a bridge where the central atom is coordinated with three adjacent atoms. Subsequently these authors refined their findings by ascribing a value of 1.5 ± 0.5 nN to rupture of a single bond within long atomic chains at cryogenic temperatures (4.2 K);[28] given that their experimental conditions favored the formation of chains 4-6 atoms long the force to rupture two and three bonds could not have been observed. From our past experiments[14, 15] and the present study, we primarily observe the formation of short single-atom bridges at room temperature where the central atom can be coordinated with one, two and three atoms on either side of the bridge. Thus, a higher value of 2.06 nN versus the previously reported 1.5 nN also suggests that shielding effects may play a stronger role in shorter bridges. Although beyond the scope of the present study and likely requiring further enhancement in sensitivity and resolution of measured forces to sub-pN level, in the future it would be of interest to explore dependence of force on chain length at low temperatures.

Finally, just as delocalization-induced resonance stabilization is the emergent response of bulk metals directional bonds and their high cohesion provides configurational stability to atomic-sized metallic devices. Moreover, similarity in the behavior of two different systems points to an inherent feature of low coordination morphologies. Results open a new opportunity for molecular electronics without complications arising from metal/molecule interfaces.

**Acknowledgments:** This work was supported by the National Science Foundation, Grant Nos. DMR-0964830, DMR-0706074, and OISE-1157130, and the support is gratefully acknowledged. HDC acknowledges useful discussions with James D. Felske. E-mails for correspondence: hchopra@buffalo.edu or zhua@buffalo.edu